\documentclass[amsmath,amssymb,onecolumn,nofootinbib,notitlepage]{revtex4-1}

 \newcommand{\be}[1]{\begin{equation}\label{#1}}
 \newcommand{\ba}[1]{\begin{eqnarray}\label{#1}}

 \newcommand{\rd}{{\rm d}}
 
 \newcommand{\re}{{\rm e}}
 \newcommand{\pa}[1]{\left(#1\right)}
 \newcommand{\paq}[1]{\left[#1\right]}
 \newcommand{\pag}[1]{\left\{#1\right\}}
 \newcommand{\av}[1]{\langle#1\rangle}
 \newcommand{\M}{{\rm M_{\rm P}}}
 \newcommand{\Mt}{{\widetilde{\rm M}_{\rm P}}}
 
 \newcommand{\mH}{{\mathcal{H}}}

 \def\ee{\end{equation}}
 \def\ea{\end{eqnarray}}

\usepackage{amsmath}

\usepackage{graphicx,latexsym}
\usepackage{graphicx,epsf}
\usepackage{color}
\usepackage{epsfig}
\usepackage{amsmath}
\usepackage[T1]{fontenc}
\usepackage[utf8]{inputenc}
\usepackage{tikz}
\usepackage{hyperref}
\usepackage[normalem]{ulem}

\begin{document}
\title{Reflected Waves and Quantum Gravity}
\author{Leonardo Chataignier}
\email{leonardo.chataignier@unibo.it}
\affiliation{Dipartimento di Fisica e Astronomia, Universit\`{a} di Bologna,
via Irnerio 46, 40126 Bologna, Italy\\I.N.F.N., Sezione di Bologna, I.S. FLAG, viale B. Pichat 6/2, 40127 Bologna, Italy}
\author{Alexander~Yu.~Kamenshchik}
\email{kamenshchik@bo.infn.it}
\affiliation{Dipartimento di Fisica e Astronomia, Universit\`{a} di Bologna,
via Irnerio 46, 40126 Bologna, Italy\\I.N.F.N., Sezione di Bologna, I.S. FLAG, viale B. Pichat 6/2, 40127 Bologna, Italy}
\author{Alessandro Tronconi}
\email{tronconi@bo.infn.it}
\affiliation{Dipartimento di Fisica e Astronomia, Universit\`{a} di Bologna,
via Irnerio 46, 40126 Bologna, Italy\\I.N.F.N., Sezione di Bologna, I.S. FLAG, viale B. Pichat 6/2, 40127 Bologna, Italy}
\author{Giovanni Venturi}
\email{giovanni.venturi@bo.infn.it}
\affiliation{Dipartimento di Fisica e Astronomia, Universit\`{a} di Bologna,
via Irnerio 46, 40126 Bologna, Italy\\I.N.F.N., Sezione di Bologna, I.S. FLAG, viale B. Pichat 6/2, 40127 Bologna, Italy}

\begin{abstract}\vspace{-0.1cm}
In the context of canonical quantum gravity, we consider the effects of a non-standard expression for the gravitational wave function on the evolution of inflationary perturbations. Such an expression and its effects may be generated by a sudden variation in the (nearly constant) inflaton potential. The resulting primordial spectra, up to the leading order, are affected in the short and in the long wavelength regime, where an oscillatory behavior with a non-negligible amplitude is superimposed on the standard semiclassical result. Moreover, a novel, non-perturbative, approach is used to study the evolution. Finally, a simplified application is fully illustrated and commented.
\end{abstract}

\maketitle\vspace{-0.5cm}
\section{Introduction}
In the last few years, the quest for the theory of quantum gravity (QG) has entered a new era. A series of increasingly precise observations, ranging from cosmic microwave background (CMB) to gravitational wave signals and the direct observation of the horizon of black holes (BHs), are now in support of the theory and may soon lead us to a consistent description of gravitational interactions at energy scales which have never been probed before and where quantum effects may be observable.

Among the several approaches to QG, canonical quantum gravity \cite{DeWitt} has an important role. It is obtained from the canonical quantization of the classical constraints emerging from the spacetime diffeomorphism invariance of general relativity. The resulting Wheeler-DeWitt (WDW) equation for the wave function of the Universe is similar to a time-independent Schr\"odinger equation in non-relativistic quantum mechanics. For simple cases, the WDW equation can be solved and a suitable interpretation of the wave function of the Universe can be given. Matter-gravity systems where the effective number of relevant degrees of freedom is small, such as BHs and inflationary cosmology, are amenable to the WDW description \cite{WDW}. Despite several conceptual issues, the canonical quantum gravity framework presents several advantages with respect to (w.r.t.) other approaches, and we expect that it is a predictive mathematical description of the QG regime, at least when quantum gravitational effects are small.

In this paper, we shall investigate some possible quantum gravitational effects in the early Universe at the energy scale of inflation \cite{inflation}, which may leave ``footprints'' in the CMB spectrum. Such effects must be small in order to fit observations but their magnitude need not be tiny and dependent on the ``usual'' $H/\M$ ratio, i.e., the ratio between the Hubble parameter and the Planck mass (see for example \cite{oscill}). As shown in \cite{BOarts}, the small (quantum) fluctuations which seed the structures we observe today can be described within this framework by a set of  separate wave functions obtained through the traditional Born-Oppenheimer (BO) decomposition \cite{BO} applied to the entire inflaton-gravity system. The approach leads to a modified Mukhanov-Sasaki (MS) equation \cite{pert} which accounts for diverse quantum gravitational effects. Non-adiabatic QG effects are obtained as a consequence of the traditional BO treatment and are tiny, being proportional to $(H/\M)^2$. Further QG effects related to the ``BO introduction of time'' can also be present \cite{time}.

Within this traditional BO scheme, the emergence of time in QG is usually associated with the ``probability current'' of the gravitational wave function. In an expanding universe, one generally assumes that the direction of such a current follows (and determines) the direction of time. However, since the gravitational wave function obeys a second-order differential equation, solutions with opposite probability fluxes always exist, and, in principle, a quantum superposition of these solutions may be considered. In any case, physical initial conditions must be imposed in order to fix the form of the gravitational wave function \cite{gravWF}, and the possibility of having a small contribution to the gravitational wave functions evolving in the opposite direction with respect to (w.r.t.) the expanding inflationary universe has been examined \cite{bounce} with the hypothesis of a bouncing universe. Here, we consider a different scenario wherein a (small) variation of the cosmological constant (inflaton potential) generates a reflected gravitational wave which influences the evolution of inflationary perturbations. The amplitude of the reflected wave will depend on the variation of the cosmological constant and its effects may be much larger than the non-adiabatic QG effects which are always present in the traditional BO approach. On a more technical level, we note that the formalism we shall employ to solve the perturbed Schr\"odinger-like equation, which governs the evolution of inflationary perturbations, is novel, and it consists in solving ``exactly'' (and numerically) the perturbed equation.

The resulting spectra are affected in the long wavelength region and in the short wavelength interval as well. In particular, in the short wavelength limit, the QG effects can be analytically well understood, as the equations can be solved with accurate approximations. In the opposite, long wavelength limit, the effects may be large but approximations are less precise. It is also important to note that we restrict our study to the case where QG effects modify the evolution of the ``CMB modes'' when such modes are still well ``inside'' the horizon. In such a case, the evolution of the perturbation modes would be insensitive to the shape of the inflaton potential, and, therefore, any variation of this potential (which is here approximated by a cosmological constant) would have no significant effect on the spectra when the QG effects derived from the WDW equation are neglected.

The article is organized as follows: In Sec. II, the general formalism is introduced; in Sec. III, the traditional BO approach is illustrated and the gravity equation is solved in the presence of a sudden variation of the cosmological constant; in Sec. IV, the matter equation in the presence of QG corrections is formally solved, and the primordial spectrum is calculated in terms of the solution of the so-called Pinney equation; in Sec. V, a simplified model is considered and its observational consequences are obtained and analyzed; finally, in Sec. VI, we draw our conclusions.

\section{Formalism}
We consider the inflaton-gravity system described by the following action
\be{fullaction}
S=\int d\eta d^3x\sqrt{-g}\paq{-\frac{\M^2}{2}R+\frac{1}{2}\partial_\mu\phi\partial^\mu\phi-V(\phi)} \,,
\ee
where $\M=\pa{8\pi G}^{-1/2}$ is the reduced Planck mass. The above action can be decomposed into a homogeneous part plus fluctuations around it. The fluctuations of the metric $\delta g_{\mu\nu}(\vec x,\eta)$ are defined as
\be{metricsc}
g_{\mu\nu}=g_{\mu\nu}^{(0)}+\delta g_{\mu\nu}\,,
\ee 
where $g_{\mu\nu}^{(0)}=\rm{diag}\paq{a(\eta)^2\pa{1,-1,-1,-1}}$ is a flat Friedmann-Lama\^{i}tre-Robertson-Walker (FLRW) metric and $\eta$ is conformal time. The scalar and the tensor fluctuations imprint their features in the CMB \cite{Planck} and are therefore the relevant perturbations during inflation. In particular, the scalar fluctuations, which can be collectively described by a single field (the MS field \cite{pert}), determine the CMB temperature fluctuations. The homogeneous degrees of freedom plus the linearized (scalar) perturbations dynamics are described by the following action
\ba{act}
S&=&\int d\eta\left\{L^3\paq{-\frac{\Mt^2}{2}a'^2+\frac{a^2}{2}\pa{\phi_0'^2-2V(\phi_{0})a^2}}\right.\nonumber\\
&+&\left.\frac{1}{2}\sum_{i=1,2}\sum_{k\neq 0}\paq{v_{i,k}'(\eta)^2+\pa{-k^2+\frac{z''}{z}}v_{i,k}(\eta)^2}\right\}\nonumber\\
&\equiv& S_G+S_I+S_{MS} \,,
\ea
where the $v_{i,k}$ are Fourier components of the scalar MS field and the index $i$ accounts for the real and imaginary parts of each component, $\Mt=\sqrt{6}\M$, $z\equiv \phi_0'/H$, $H=a'/a^2$ is the Hubble parameter, the prime denotes the derivative w.r.t. conformal time and 
\be{defL}
L^3\equiv \int d^3x.
\ee
Notice that we formally split the full action into three contributions: $S_G$ and $S_I$ are the homogeneous gravity and inflaton actions, respectively, whereas $S_{MS}$ describes the perturbations.

Henceforth, we shall set $L$ to be equal to $1$ (see \cite{BOarts} for more details) to keep the notation compact. The Hamiltonian is finally
\begin{eqnarray}
&&\mathcal{H}=-\frac{\pi_a^2}{2\Mt^2}+\pa{\frac{\pi_\phi^2}{2a^2}+a^4 V}+\sum_{k\neq 0}^{\infty}\paq{\frac{\pi_k^2}{2}+\frac{\omega_k^2}{2}v_k^2} \,,
\label{ham}
\end{eqnarray}
where $\omega_k^2=k^2-z''/z$. For simplicity, we shall limit ourselves to the case of a constant inflaton potential $V=\Lambda$. One has
\be{momenta}
\pi_a=-\Mt^2 a'\,,\;\pi_\phi=a^2\phi_0'\,,\;\pi_k=v'_k.
\ee
The canonical quantization of the Hamiltonian constraint (\ref{ham}) leads to the following WDW equation for the wave function of the Universe (matter plus gravity)
\be{WDW0}
\left\{\frac{1}{2\Mt ^2}\frac{\partial^2}{\partial a^2}+{\hat\mH}_0+\sum_{k\neq 0}^{\infty}\hat {\mathcal H}_k\right\}\Psi\pa{a,\phi_0,\{v_k\}}=0 \,,
\ee
where we define
\begin{align}
{\hat\mH}_0 &:= -\frac{1}{2a^2}\frac{\partial^2}{\partial \phi_0^2}+a^4\Lambda \ , \\
\hat {\mathcal H}_k&:= \frac{1}{2}\frac{\partial^2}{\partial v_k^2}+\frac{\omega_k^2}{2}v_k^2 \ ,
\end{align}
for $k\neq 0$. Equation (\ref{WDW0}) will be the starting point in our approach. 

\section{BO decomposition}
In this section, we briefly illustrate the traditional BO decomposition for the inflaton-gravity system. This decomposition is one of the approaches to the problem of time in quantum gravity, i.e., to the introduction of an evolution parameter in the seemingly stationary equation (\ref{WDW0}), and it has a rather long history (see \cite{chataig:rev} for a review). In its traditional version, the BO decomposition follows the formalism described, e.g., in \cite{Hunter}, where a factorization of the wave function leads to the decomposition of the time-independent Schr\"{o}dinger equation [here, given by (\ref{WDW0})] into an equation for ``heavy'' variables (here, the scale factor) and one for the ``light'' variables (here, the matter fields), and both equations are, in principle, nonlinear (as they include so-called non-adiabatic effects). The key idea of the traditional BO decomposition, as applied to quantum cosmology, is that the phase of the gravitational wave function can be used to define a time variable that dictates the evolution of matter fields. The nonlinearities of the matter equation can be dealt with by suitable ``re-phasings'' of the gravitational and matter wave functions (see \cite{chataig:rev,unitarity,equivalence}) and by a concrete, iterative procedure that uses perturbation theory (usually in powers of the inverse Planck mass, see \cite{BOarts,equivalence,beyond,otherBO} and references therein for further details; see also \cite{hybrid} for an application of a BO-inspired approach in ``hybrid quantum cosmology'' with techniques used in loop quantum cosmology, and \cite{unitarity} for a comparison of diverse approaches). In the spirit of this traditional BO approach, we thus start from the ansatz
\be{tradBO}
\Psi\pa{a,\phi_0,\pag{v_k}}=\psi(a)\chi_M(a,\phi_0,\pag{v_k})=\psi(a)\chi_0(a,\phi)\prod_{k\neq 0}\chi_k(a,v_k).
\ee
If we project out the matter wave function, one is then led to a gravity equation
\be{greq}
\partial_a^2\tilde \psi+2\Mt^2\av{\hat {\mathcal H}_0}\tilde \psi
=\av{\partial_a\tilde \chi_0|\partial_a \tilde \chi_0}\tilde \psi,
\ee
where
\begin{align}
\psi &=\tilde{\psi}\, \re^{-\mathrm{i}\sum_k\int^a A_k\rd a'} \,,\\
\chi_k &= \tilde{\chi}_k\,\re^{\mathrm{i}\int^a A_k\rd a'} \,,\\
A_k &:= -\mathrm{i}\av{\chi_k|\partial_a|\chi_k}\,,\\
\av{\hat O}&=\int_{-\infty}^{+\infty}\pa{\prod_{k\neq 0}\rd v_k}\rd \phi\,\chi_M^*\hat O\chi_M \,,\label{defO}
\end{align}
with $k=0$ indicating the homogeneous scalar field. Let us note that we neglected the back-reaction originating from the perturbations $v_k$ in (\ref{greq}).

The projection of the WDW along $\prod_{p\neq k}\langle \chi_{p}|$ leads to a set of equations, one for each $k$-mode of the inflaton, of the following form
\be{mateq}
\frac{1}{\Mt^2}\frac{\partial_a\tilde \psi}{\tilde \psi}\partial_a\tilde \chi_k+\pa{\hat {\mathcal H}_k-\av{\hat{\mathcal H}_k}_k}\tilde \chi_k+\frac{1}{2\Mt^2}\pa{\partial_a^2-\av{\tilde \partial_a^2}_k}\tilde \chi_k=0\,,
\ee
where 
\be{deftildeO}
\av{\hat {\tilde O}}_k=\int_{-\infty}^{+\infty}\rd v_k \tilde\chi_k^*\hat O\tilde \chi_k.
\ee
{We note that (\ref{mateq}) plays a central role in our approach to the fully quantized inflaton-gravity system. Its structure is a consequence of the BO decomposition performed and it contains: a first term with the first derivative of the matter wave function and also dependent on gravitational wave function (this term is usually associated with the introduction of time), the Hamiltonian contribution given by the matter Hamiltonian minus its expectation value, and a third term given by the second derivative of matter wave function minus its expectation value. This latter term is associated with the non-adiabatic effects in the traditional BO decomposition (when such a BO decomposition is applied to molecules it describes the effects of the ``slowly'' moving nucleus on the electron cloud) and in this context it is usually associated with quantum gravitational effects. Let us note that also the first contribution may give rise to quantum gravitational effects associated with the introduction/definition of time and we then keep these letter effects distinct w.r.t. to the former ones. Let us further note, as discussed in detail in \cite{unitarity}, that, the structure of (\ref{mateq}), apart from the first term, has the form $\av{\hat{ \tilde O}}_k-\hat O$, where $\hat O$ is not necessarily Hermitian. Thus, since $\Mt^{-2}\partial_a\tilde \psi/\tilde \psi$ appears as a c-number, one has
\be{propr1}
\langle \tilde \chi_k|\partial_a|\tilde \chi_k\rangle=0 \,,
\ee
and
\be{prop2}
\partial_a\langle \tilde \chi_k|\tilde \chi_k\rangle=\langle \partial_a\tilde \chi_k|\tilde \chi_k\rangle+\langle\tilde \chi_k| \partial_a\tilde \chi_k\rangle=\frac{\Mt^2\tilde \psi^*}{\partial_a\tilde \psi^*}\pa{\langle \tilde \chi_k|\hat O^\dagger-\av{\hat {\tilde O}^\dagger}_k|\tilde \chi_k\rangle}+\frac{\Mt^2\tilde \psi}{\partial_a\tilde \psi}\pa{\langle \tilde \chi_k|\hat O-\av{\hat {\tilde O}}_k|\tilde \chi_k\rangle}=0
\ee
which means $\langle \tilde \chi_k|\tilde \chi_k\rangle$ normalization is conserved w.r.t. the variation of $a$ or of any function of it (and in particular the semiclassical time).}

If $k=0$, the equation (\ref{mateq}) is that of the homogeneous inflaton. When the inflaton potential is constant, such an equation simplifies to 
\be{mateq0}
\frac{1}{\Mt^2}\frac{\partial_a\tilde \psi}{\tilde \psi}\partial_a\tilde \chi_0-\frac{1}{2a^2}\pa{\frac{\partial^2}{\partial\phi_0^2}-\av{\frac{\partial^2}{\partial\phi_0^2}}_0}\tilde \chi_0+\frac{1}{2\Mt^2}\pa{\partial_a^2-\av{\tilde \partial_a^2}_0}\tilde \chi_0=0 \,,
\ee
and it is easily solved by (scale-factor independent) eigenstates of the operator $-i\partial/\partial\phi_0$ (i.e., plane waves $\tilde \chi_0\sim \re^{i P \phi_0/\Mt}$). Correspondingly, the gravity equation (\ref{greq}) becomes
\be{greq1}
\partial_a^2\tilde \psi+2\Mt^2\pa{\frac{P^2}{2a^2\Mt^2}+a^4\Lambda}\tilde \psi=0\,,
\ee
and it can be solved as well. In particular, its solutions have a simple form in the large $a$ limit. Let $y=\Mt^3a^3$ and $\Lambda\equiv \Mt^4 \lambda$, then (\ref{greq1}) becomes
\be{greqy}
\partial_y^2\tilde \psi+\frac{2}{3y}\partial_y\tilde \psi+\pa{\frac{P^2}{9y^2}+\frac{2}{9}\lambda}\tilde \psi=0,
\ee
and for $y\gg 1$ its solution is a (quantum and coherent) superposition of plane waves moving forward and backward in ``time'', respectively
\be{solgr}
\tilde\psi=A_1\re^{-i\sqrt{2\lambda} y/3}+A_2\re^{i\sqrt{2\lambda} y/3}.
\ee 
Notice that the expression becomes increasingly accurate in the $y\gg 1$ limit provided $P/y\rightarrow 0$ in the same limit.
In particular, for (\ref{solgr}), the second term of Eq. (\ref{greqy}) is $\sim \sqrt{\lambda}/y\,\tilde \psi$ and is much smaller than the remaining two if
\be{compapp}
\frac{\sqrt{\lambda}}{\Mt^3 a^3}\ll \lambda\Rightarrow a\gg \frac{1}{\Mt\lambda^{1/6}}.
\ee
The matter equation (\ref{mateq}) has three contributions: the first, which contains the gravitational wave function, is associated with the introduction of time; the Hamiltonian determines the evolution of the matter wave function; and the third contribution describes the non-adiabatic corrections and is generally proportional to $H^2/\Mt^2$. During inflation, when $H/\Mt\ll 1$, this last contribution is generally tiny. Still, it is a possible source of quantum gravitational effects, and its contribution has been studied in several articles \cite{BOarts}.

Let us now suppose that $\tilde\psi\propto\re^{-i\sqrt{2\lambda} y/3}$. Then, the first term in the matter equation is
\be{time0}
\frac{1}{\Mt^2}(\partial_a y)\frac{\partial_y \tilde\psi}{\tilde\psi}\partial_a \tilde \chi_k=-i a^2\sqrt{ \frac{\Lambda}{3\M^2}}\,\partial_a\tilde \chi_k=-i a'\partial_a \tilde \chi_k\equiv-i\frac{\rd\tilde\chi_k}{\rd \eta} \,,
\ee
where the classical Friedmann equation has been used to relate the cosmological constant to the classical velocity $a'$, and the ``classical'' time can then be introduced in the matter equation. In the traditional BO approach, the ``flow'' of time is related to the phase of the gravitational wave function. For $a\Mt$ small, the gravitational wave function has a behavior that is very different from a simple plane wave, and time cannot be introduced. Moreover, in the regime where such a difference is small, time can still be introduced but quantum gravitational fluctuations are also present and may have observable consequences. These fluctuations may be much larger than those associated with the non-adiabatic effects. In this article we shall consider some of their possible origins and their resulting effects. In particular, we study how small variations in the (nearly constant) inflaton potential may affect the gravitational wave function and the resulting MS equation. Small variations of the inflaton potential modify the propagation of the MS field $v_k$ after the corresponding momenta exit the horizon ($k/(aH)\sim 1$) and remains imprinted in the CMB. However, the MS equation is nearly insensitive to such variations in the limit $k/(aH)\gg 1$ when essentially $v_k$ evolves as a plane wave. Nevertheless, in this case, the QG effects originated by a modification of the gravitational wave function can be present. In previous treatments of the traditional BO approach, such effects are not present, but they can be studied within our current approach. For simplicity, the variation of the inflaton potential will be modeled by sudden changes of cosmological constant occurring at certain time (and, correspondingly, certain values of the scale factor). 

\subsection{Potential Well}
Let us now consider a small variation of the cosmological constant $\Delta \Lambda\equiv \Mt^4\,\Delta \lambda$ in the interval $[y_1,y_2]$ with $y_{1,2}\gg 1$. In this case, one must solve the gravitational equation (\ref{greqy}) in 3 regions: the incoming region $\rm{(I)}$ with $y<y_1$, the region $\rm{(II)}$ inside the interval, and the outgoing region $\rm{(III)}$ with $y>y_2$. In the latter region only an outgoing wave must be present and, therefore, the resulting wave functions are:
\be{r1}
\tilde\psi_{\rm I}=\re^{-i q y}+r\re^{i q y},\quad \tilde\psi_{\rm{II}}=a\re^{i p y}+b\re^{-i p y},\quad \tilde\psi_{\rm{III}}=t\re^{-i q y}\,,
\ee
with $q=\sqrt{2\lambda/9}$, $p=\sqrt{2\pa{\lambda+\Delta \lambda}/9}$ and $\Delta \lambda$ may be a positive or negative (small) variation of $\lambda$.
On imposing the junction conditions and defining $y_2\equiv y_1+\Delta y$, one has
\be{t}
t=\frac{2q\,p\,\re^{i q\,\Delta y}}{2q\,p\cos\pa{p\,\Delta y}+i\pa{q^2+p^2}\sin\pa{p\,\Delta y}},
\ee

\be{ab}
a=\frac{(p-q)\re^{-i\pa{q+p}y_2}}{2\,p\,}t,\; b=\frac{(q+p)\re^{-i\pa{q-p}y_2}}{2\,p}t,
\ee

\be{r}
r=\frac{i(q^2-p^2)\sin\pa{p\,\Delta y}\re^{-iq\pa{y_1+y_2}}}{2q\,p}t.
\ee
The condition $|t|=1$ is equivalent to the requirement $p\,\Delta y=n\pi$ and $r=0$ and is known as the ``resonant transmission'' condition. In this case, a reflected gravitational wave function is only present inside the $[y_1,y_2]$ interval, inside which the time flow is modified. The ratio
\be{aob}
\frac{a}{b}=\frac{p-q}{p+q}\re^{-2i p\,y_2}\equiv \epsilon\,\re^{-2i p\,y_2}
\ee
determines the relative weight of the ingoing and outgoing wave and its modulus must be small in order to properly define the ``classical time flow''. The resulting, total wave function in the ``resonant transmission'' case is
\be{totWF}
\psi=\re^{-i q y}\paq{\theta\!\pa{y_1-y}+\theta\!\pa{y-y_2}}+b\,\re^{-i p y}\pa{1+\epsilon\,\re^{2 i p \pa{y-y_1}}}\theta\!\pa{y-y_1}\theta\!\pa{y_2-y}\,,
\ee
where $\theta(x)$ is the Heaviside theta function and we used the condition $p\,\Delta y=n\pi$ to express the exponent in terms of $y_1$. One then has to account for the reflected wave only in a certain time interval, outside which the time ``flow'' mimics the semiclassical evolution, and no QG effects (except the tiny non-adiabatic contributions) are present. The ``resonant transmission'' case, despite being less general, has the advantage of restricting the QG effects to a given period of the inflationary evolution, and it does not affect the initial state  [Bunch-Davies (BD) vacuum \cite{BD}] of the inflationary perturbations. In contrast, the presence of a reflected wave in the interval $y<y_1$ would require a proper reexamination of the definition of the initial condition for the MS field. Therefore, we restrict our attention to the ``resonant transmission'' case, and avoid the complications of more general setups.

\subsection{Potential barrier}
A slightly simplified setup w.r.t that described above is illustrated in what follows. Let us suppose that a small variation of the cosmological constant is present at some value of the scale factor $y_0$. Then, one must solve the gravitational equation (\ref{greqy}) in 2 regions: the incoming region $\rm{(in)}$ with $y<y_0$ and the outgoing region $\rm{(out)}$ with $y>y_0$
\be{r0}
\tilde\psi_{\rm{in}}=\re^{-i p y}+r\re^{i p y},\quad \tilde\psi_{\rm{out}}=t\re^{-i q y} \,.
\ee
On imposing the junction conditions one finds:
\be{rt}
r=\frac{p-q}{p+q}\re^{-2 i p y_0},\;t=\frac{2p}{p+q}\re^{i(q-p)y_0}\,,
\ee
and $|r|$ must be small in order to define a ``classical time flow'' and a (quantum gravitational) perturbation originated by the reflected wave in the $\rm{(in)}$ region. The smallness of $|r|$ is equivalent to the condition 
\be{rsmall}
\frac{p-q}{p+q}\equiv \epsilon,\;|\epsilon|\ll 1.
\ee
The resulting, total wave function is
\be{totWF}
\psi=\re^{-i p y}\pa{1+\epsilon\, \re^{2ip\pa{y-y_0}}}\theta\!\pa{y_0-y}+t\re^{-i q y}\theta\!\pa{y-y_0}.
\ee 
As already discussed previously, the presence of a reflected wave for $y<y_0$ necessarily leads to a problematic discussion on initial conditions (of the MS field) and we shall not consider this case in the final analysis.\footnote{\label{foot:model}It is worthwhile to mention that the two cases considered here are relatively simple: one in which there is just one barrier, and hence a reflected wave; and the other (resonant transmission) for which there is no reflection of the incoming wave. These by themselves generate observable effects. Further refinements in the model can, of course, be contemplated, but similar results to our current setup are expected, e.g., for the nonresonant case. However, the presence of a reflected wave in the first interval would lead to more complicated calculations and heavier numerical simulations.}  
\\
\section{Mukhanov-Sasaki equation}
If we neglect the non-adiabatic contributions (which are tiny during inflation), the matter equation (\ref{mateq}) simplifies to 
\be{pMS}
\frac{1}{\Mt^2}\frac{\partial_a\tilde \psi}{\tilde \psi}\partial_a\tilde \chi_k+\pa{\hat {\mathcal H}_k-\av{\hat{\mathcal H}_k}}\tilde \chi_k=0\,,
\ee
{where henceforth the subscript $k$ in the expectation values in the MS equation will be omitted in order to keep the notation compact.} When the gravitational wave function is a ``simple'' outgoing wave, apart from an overall phase redefinition, equation (\ref{pMS}) reduces to the Schr\"odinger representation of the standard MS equation. The presence of a small ingoing (reflected) contribution still allows one to properly define the classical time flow but one must also account for a small quantum gravitation perturbation around it. In this case, the perturbed gravitational wave function has the general form
\be{pgwf}
\tilde \psi\propto\re^{-i p y}\pa{1+\epsilon\, \re^{2ip\pa{y-y_0}}}\,,
\ee
and $y_0\rightarrow y_2$ in the resonant transmission case, where $\epsilon$ is a real constant. The first term in (\ref{pMS}) becomes 
\be{dpsi}
\frac{1}{\Mt^2}\frac{\partial_a\tilde \psi}{\tilde \psi}\partial_a=-3ip\Mt a^2\pa{1-\frac{2\,\epsilon\,  \re^{2ip\pa{y-y_0}}}{1+\epsilon \,\re^{2ip\pa{y-y_0}}}}\partial_a\equiv -i\frac{p}{q}\pa{\frac{1-\,\epsilon\,  \re^{2ip\pa{y-y_0}}}{1+\epsilon \,\re^{2ip\pa{y-y_0}}}}\frac{\rd}{\rd \eta}\,,
\ee
and it is continuous on varying $y$ due to the junction conditions imposed on the gravitational wave function. However, we observe that $\partial_a(\partial_a\tilde \psi/\tilde \psi)$ is necessarily discontinuous due to the discontinuity of the source term (cosmological constant) in the gravity equation (\ref{greq1}). A more realistic scenario would include a smooth variation of a slowly varying inflaton potential (in fact, of all of the physical parameters and their derivatives) instead of a sudden jump of its energy density. This more realistic analysis would lead to equations that are considerably more complicated without essentially adding relevant details to the overall physical discussion. 

In Eq. (\ref{dpsi}), the conformal time variable was introduced by the following change of variable $y=\Mt^3a^3=-\Mt^3/(H^3 \eta^3)$ (which is valid in de Sitter) and it is physically related to the probability current of the leading (outgoing) term of the gravitational wave function (\ref{pgwf}). The perturbed MS equation then becomes
\be{pms}
i\frac{\rd\tilde \chi_k}{\rd \eta}=\frac{1-\epsilon}{1+\epsilon}\cdot\frac{1+\,\epsilon\,  \re^{2ip\pa{y-y_0}}}{1-\,\epsilon\,  \re^{2ip\pa{y-y_0}}}\pa{\hat {\mathcal H}_k-\av{\hat{\mathcal H}_k}}\tilde \chi_k\,,
\ee
where (\ref{aob}) has been used to express $p/q$ in terms of $\epsilon$. In spite of the explicit non-Hermitian form of Eq. (\ref{pms}), one has 
\be{uni}
\frac{\rd}{\rd \eta}\langle \tilde \chi_k|\tilde \chi_k\rangle=i\frac{1-\epsilon}{1+\epsilon}\cdot\frac{1+\,\epsilon\,  \re^{2ip\pa{y-y_0}}}{1-\,\epsilon\,  \re^{2ip\pa{y-y_0}}}\pa{\av{\hat{\mathcal H}_k}-\av{\hat{\mathcal H}_k}}+{\rm h.c.}=0\,;
\ee
i.e., the norm of $\tilde \chi_k$ is conserved and can be arbitrarily set to one. Moreover $\langle \tilde \chi_k|\rd/\rd \eta|\tilde \chi_k\rangle=0$ {[as expected given the general properties discussed in \cite{unitarity} and reviewed in (\ref{propr1}) and (\ref{prop2})]}.

{Instead of directly solving the modified MS equation (\ref{pms}) it is convenient to define a new function by} $\tilde\chi_k=\re^{i\varphi}\chi_k$, where $\varphi$ is a complex function of $\eta$ given by
\be{varphi}
\frac{\rd \varphi}{\rd \eta}=\frac{1-\epsilon}{1+\epsilon}\cdot\frac{1+\,\epsilon\,  \re^{2ip\pa{y-y_0}}}{1-\,\epsilon\,  \re^{2ip\pa{y-y_0}}}\av{\hat{\mathcal H}_k}\,,
\ee
and, finally, obtain
\be{newMS}
i\frac{\rd}{\rd \eta}\chi_k=\frac{1-\epsilon}{1+\epsilon}\cdot\frac{1+\,\epsilon\,  \re^{2ip\pa{y-y_0}}}{1-\,\epsilon\,  \re^{2ip\pa{y-y_0}}}\hat {\mathcal H}_k \chi_k\equiv \hat{\mathcal  \bar H}_k \chi_k
\ee
{having the conventional form of the Schr\"odinger equation.} 
In the ``resonant transmission'' case, if we define 
\be{mdef}
m=\paq{\theta\!\pa{y_1-y}+\theta\!\pa{y-y_2}}+\theta\!\pa{y-y_1}\theta\!\pa{y_2-y}\frac{1+\epsilon}{1-\epsilon}\cdot\frac{1-\,\epsilon\,  \re^{2ip\pa{y-y_1}}}{1+\,\epsilon\,  \re^{2ip\pa{y-y_1}}}\,,
\ee 
the MS hamiltonian transforms into
\be{pertH}
 \hat{\mathcal  H}_k=\frac{1}{2}\hat \pi_v^2+\frac{\omega^2}{2}\hat v^2\rightarrow  \hat{\mathcal  \bar H}_k= \frac{1}{2m}\hat \pi_v^2+\frac{m\tilde\omega^2}{2}\hat v^2\,,
\ee
with a time-dependent complex mass and the new frequency $\tilde \omega^2\equiv \omega^2/m^2$. {We note that the definition of $\chi_k$ is a matter of convenience since in terms of $\chi_k$ the equation (\ref{pms}) takes explicitly the form of the Schr\"odinger equation for a time dependent, non-Hermitian harmonic oscillator and can be formally solved by the adiabatic invariant technique \cite{invariant}. Moreover $\rd/\rd \eta \langle \chi_k|\chi_k\rangle\neq 0$, as it should for a solution of a non-Hermitian Schr\"odinger equation. Once the formal solution of (\ref{newMS}) is obtained, one can transform it back and obtain the correctly normalized solution of (\ref{pms}).}

It is important to note that one recovers the standard MS equation in the $\epsilon\rightarrow 0$ limit. Moreover, $m(y)$ is continuous in $y$ and is $1$ for $y<y_1$ and $y>y_2$ but its derivative is discontinuous. As we already mentioned, the step functions would be replaced by smooth functions and the derivatives of $m(y)$ would be continuous in a more realistic scenario. Let us finally note that henceforth, since the notion of time has been introduced, we shall adopt $\eta_{1}$, and $\eta_2$ instead of the corresponding values of $y_1$ and $y_2$. 

\subsection{Invariant operators}
Formally, the solution for the non-Hermitian harmonic oscillator is identical to its Hermitian counterpart {(see for example \cite{Zelaya})}. One can define the following quadratic invariant operator
\be{lininv}
\hat I=\frac{1}{2}\paq{\pa{\frac{\hat v}{\rho}}^2+\pa{\rho \hat \pi_v-m\rho'\hat v}^2}\,,
\ee
and the invariant vacuum state is its zeroth order eigenstate, which is given by
\be{vac}
\chi_{k,0}=\pa{\frac{1}{\pi\rho^2}}^{1/4}\exp\paq{i\frac{m}{2}\pa{\frac{\rho'}{\rho}+\frac{i}{m\rho^2}}v^2-\frac{i}{2}\int_{\eta_i}^{\eta}\frac{\rd \bar\eta}{m\rho^2}}\,.
\ee
In (\ref{vac}), $\rho$ is the Pinney variable. It satisfies the following second order, non-linear ``Pinney'' ODE
\be{pinney}
\rho''+\frac{m'}{m}\rho'+\tilde\omega^2\rho=\frac{1}{m^2\rho^3}.
\ee
If we employ the technique of invariants, we can obtain the solution of a partial differential equation (PDE) -- the Schr\"odinger equation -- from the solution of a non-linear ODE (the Pinney equation), which can be solved numerically in a straightforward way. Notice that (\ref{vac}) is the invariant vacuum solution {of (\ref{newMS}), as can be easily checked by direct substitution}, and it is valid regardless of whether quantum gravitational corrections (ingoing wave) are present. Technically, in order to find the solution of the MS vacuum wave function (\ref{vac}), we must solve the Pinney equation in the unperturbed and in the perturbed regions separately, and subsequently set the initial conditions in each region by appropriate junction conditions. Therefore, each region is characterized by its own initial conditions for the Pinney variable. This guarantees the continuity of $\rho$ and its first derivative which, as we shall see, are the quantities which determine the main observable of the model: the power spectrum.

When one solves the MS equation in the standard semiclassical case, the initial conditions are set at $\eta_i\rightarrow -\infty$, and the BD vacuum is usually taken to be the correct initial state of the inflationary perturbations. However, the limit in which $|\eta|$ is large corresponds to the realm of QG, in the $a\Mt\le \lambda^{-1/6}$ region, where time ``does not exist'' (according to the usual view employed in the traditional BO approach -- see Eq. (\ref{compapp})) and the semiclassical formalism is not applicable. We therefore assume that there exists some finite time, $\eta_P=-1/(H a_P)$, at which $a_P\Mt\gg \lambda^{-1/6}$ and ``conservative'' initial conditions are imposed. On the de Sitter background, the exact solution of the Pinney equation that corresponds to the BD vacuum state is
\be{rhods}
\rho_{\rm dS}(\eta)=\sqrt{\frac{1+k^2\eta^2}{k^3\eta^2}},\; \rho'_{\rm dS}(\eta)=\frac{1}{k^{3/2}\eta^2\sqrt{1+k^2\eta^2}}\,,
\ee
and we assume $\rho(\eta_P)=\rho_{\rm dS}(\eta_P)$ and $\rho'(\eta_P)=\rho'_{\rm dS}(\eta_P)$. Although the initial conditions corresponding to the BD vacuum are real, the Pinney equation (\ref{pinney}) and its solution are complex in the presence of the quantum gravitational effects we are analyzing.\footnote{Although we fix initial conditions that correspond to the usual BD vacuum, it is clear that more complicated choices can also be made (see, for instance, \cite{oscill}) but they, as in our case, will lead to structures in the CMB spectrum. Our approach is, however, quite distinct, since effects are introduced solely through the gravitational wave function and the sudden variation in the nearly constant inflaton potential. See also footnote \ref{foot:model}.} Therefore, if these effects are present, $\rho$ becomes complex and the vacuum wave function (\ref{vac}) has the general form
\be{vaccomplex}
\chi_{k,0}=A_0(\eta)\re^{i\gamma_0(\eta)}\exp\paq{i(\alpha_0(\eta)+i \beta_0(\eta))\hat v^2}\,,
\ee
where $A_0$, $\alpha_0$, $\beta_0$ and $\gamma_0$ are complex functions of time and
\be{deffun01}
A_0(\eta)\re^{i\gamma_0(\eta)}\equiv \pa{\frac{1}{\pi \rho^2}}\re^{-\frac{i}{2}\int_{\eta_1}^{\eta}\frac{\rd \bar\eta}{m\rho^2}},\ee
\be{deffun02}
\alpha_0={\rm Re}\paq{\frac{1}{2}\pa{\frac{m\rho'}{\rho}+\frac{i}{\rho^2}}},\;\beta_0={\rm Im}\paq{\frac{1}{2}\pa{\frac{m\rho'}{\rho}+\frac{i}{\rho^2}}}.
\ee
This general complex form is maintained when the quantum gravitational effects disappear (and $m=1$).

\subsection{Power Spectrum}
We are interested in calculating the quantum gravitational effects generated by the interference between the ingoing and the outgoing gravitational wave function on the power spectrum. The power spectrum is defined {in term of the physical matter state $\tilde\chi_k$} as
\be{PSdef}
\Delta_s^2=\lim_{-k \eta\rightarrow 0}\frac{k^3}{2\pi^2}\langle\tilde\chi_{k,0}|\hat v^2|\tilde \chi_{k,0}\rangle\,,
\ee
where $|\tilde \chi_{k,0}\rangle$ is the vacuum and its normalization is constant [see (\ref{uni})]:
\be{normchi}
\langle\tilde\chi_{k,0}|\tilde \chi_{k,0}\rangle=1=\langle\chi_{k,0}| \chi_{k,0}\rangle \re^{2i\,{\rm Im}\varphi}\Rightarrow \re^{2i\,{\rm Im}\varphi}=\langle\chi_{k,0}| \chi_{k,0}\rangle^{-1}.
\ee
{The power spectrum (\ref{PSdef}) can be rewritten} in terms of $|\chi_{k,0}\rangle$, given by the expression (\ref{vac}), and we have
\be{PSdef2}
\Delta_s^2=\lim_{-k \eta\rightarrow 0}\frac{k^3}{2\pi^2}\langle\chi_{k,0}|\hat v^2| \chi_{k,0}\rangle \re^{2i\,{\rm Im}\varphi}=\lim_{-k \eta\rightarrow 0}\frac{k^3}{2\pi^2}\frac{\langle\chi_{k,0}|\hat v^2| \chi_{k,0}\rangle}{\langle\chi_{k,0}| \chi_{k,0}\rangle}\,,
\ee
as it should be for a non-normalized wave function. Finally, from (\ref{vaccomplex}), one finds
\be{avv2}
\frac{\langle\chi_{k,0}|\hat v^2| \chi_{k,0}\rangle}{\langle\chi_{k,0}| \chi_{k,0}\rangle}=\frac{\int_{-\infty}^{+\infty}v^2 \re^{-2\beta_0 v^2}}{\int_{-\infty}^{+\infty}\re^{-2\beta_0 v^2}}=\frac{1}{4\beta_0}\,,
\ee
where time dependent normalization factors and complex phases are canceled. The integral (\ref{avv2}) only converges for $\beta_0>0$, which must be verified a posteriori. One is usually interested in the power spectrum at the end of inflation ($\eta\rightarrow 0^-$), when the modes are well outside the horizon ($\lim_{-k \eta\rightarrow 0}$). In this limit, $m=1$ and 
\be{finalPS}
\Delta_s^2=\lim_{-k \eta\rightarrow 0}\frac{k^3}{2\pi^2}{\rm Im}^{-1}\paq{2\pa{\frac{\rho'}{\rho}+\frac{i}{\rho^2}}}.
\ee
The Pinney variable in (\ref{finalPS}) can be calculated analytically for $\eta>\eta_2$ once $\rho_0\equiv\rho(\eta_2)$, $\rho'_0\equiv\rho'(\eta_2)$ (for the ``resonant transmission'' case) are known. Indeed, the QG effects are absent ($m=1$) at the end of inflation, and the Pinney equation can be solved exactly. The difference w.r.t. the standard de Sitter case is comprised only of the initial conditions.

When $\eta>\eta_2$, the Pinney equation takes the standard de Sitter form with $m=1$ and $m'=0$. Its general solution can then be written in terms of two independent solutions of the associated homogeneous equation,
\be{hompin}
x''+\omega^2 x=0\,.
\ee
For example,
\be{indepsol}
x_{1,2}(\eta)=\pa{1\pm\frac{i}{k\eta}}\re^{\pm ik\eta}\stackrel{-k\eta\rightarrow 0^+}{\longrightarrow}\pm\frac{i}{k\eta}
\ee
and, correspondingly,
\be{dindepsol}
x_{1,2}'(\eta)=\mp i\frac{1\mp i\,k\eta-k^2\eta^2}{k\eta^2}\re^{\pm ik\eta}\stackrel{-k\eta\rightarrow 0^+}{\longrightarrow}\mp\frac{i}{k\eta^2}=\mp\frac{x_{1,2}(\eta)}{\eta}\,.
\ee
Let $u$ and $v$ be two independent solutions of (\ref{hompin}). Their Wronskian $W=u'v-uv'$ is constant and the expression
\be{rhosol}
\rho=\sqrt{u^2+\frac{v^2}{W^2}}
\ee
is a solution to $\rho''+\omega^2\rho-1/\rho^3=0$. In order to reproduce the desired initial conditions $\rho_0$ and $\rho'_0$ for the associated Pinney equation at $\eta=\eta_2$, it is convenient to choose, if possible, $v(\eta_2)=0$ and $v'(\eta_2)\neq 0$ and arbitrary. Then, if $\rho_0\neq 0$, one has
\be{inipin}
\rho_0=u(\eta_2),\; \rho'_0=u'(\eta_2)\,;
\ee
i.e., the initial conditions for $\rho(\eta)$ and $u(\eta)$ coincide. In terms of (\ref{indepsol}), one obtains
\be{v=0}
v(\eta)=x_2(\eta)-\frac{\pa{k\eta_2-i}^2}{1+k^2\eta_2^2}\re^{-2ik\eta_2}x_1(\eta)\stackrel{-k\eta\rightarrow 0^+}{\longrightarrow}-\frac{i}{k\eta}\pa{1+\frac{\pa{k\eta_2-i}^2}{1+k^2\eta_2^2}\re^{-2ik\eta_2}}\,,
\ee
and
\be{urho}
u(\eta)=A_1 x_1(\eta)+A_2 x_2(\eta)\stackrel{-k\eta\rightarrow 0^+}{\longrightarrow}\frac{i}{k\eta}(A_1-A_2)\,,
\ee
where
\be{A1}
A_1=\frac{\re^{-ik\eta_2}}{2k^2\eta_2^2}\paq{\rho_0\pa{k^2\eta_2^2-ik\eta_2-1}-\eta_2\rho'_0\pa{ik\eta_2+1}}\,,
\ee
and
\be{A2}
A_2=\frac{\re^{ik\eta_2}}{2k^2\eta_2^2}\paq{\rho_0\pa{k^2\eta_2^2+ik\eta_2-1}-\eta_2\rho'_0\pa{-ik\eta_2+1}}.
\ee
The parameters $A_2=A_1^*$ if $\rho_0$ and $\rho'_0$ are real numbers. Notice that
\be{dror}
\frac{\rho'}{\rho}=\frac{uu'+\frac{vv'}{W^2}}{u^2+\frac{v^2}{W^2}}\stackrel{-k\eta\rightarrow 0^+}{\longrightarrow}-\frac{1}{\eta}\,,
\ee
because $v'=-v/\eta$ and $u'=-u/\eta$ in the long wavelength limit. Therefore, the expression (\ref{finalPS}) simplifies to
\be{finalPSsim}
\Delta_s^2=\lim_{\eta\rightarrow 0^-}\frac{k^3}{2\pi^2}{\rm Re}\frac{\rho^2}{2}\,,
\ee
which reduces to the standard de Sitter result when the QG effects are absent. For the de Sitter case, we find 
\be{PS2}
\rho_{dS}^2\stackrel{\eta\rightarrow 0^-}{\longrightarrow}\frac{1}{k^3\eta^2}\,.
\ee
It is then convenient to introduce (and plot) the quantity 
\be{finalPSsimtilde}
\widetilde{\Delta}_s^2=\lim_{\eta\rightarrow 0^-}\frac{k^3}{2\pi^2}{\rm Re}\frac{\eta^2\rho^2}{2}\,,
\ee
which is not divergent in the $\eta\rightarrow 0^-$ limit.

For the modes inside the horizon at $\eta_2$ ($-k\eta_2\gg 1$), the above expressions can be rewritten as follows:
\be{uvin}
v_{\rm in}=-\frac{i}{k\eta}\pa{1+\re^{-2ik\eta_2}}, \;u_{\rm in}=\frac{1}{k\eta}\pa{\rho_0\sin k\eta_2+\frac{\rho_0'}{k}\cos k\eta_2}\,,
\ee
where $\rho_0$ and $\rho_0'$ are complex numbers, and, finally,
\be{rho2in}
\rho^2_{\rm in}=\frac{1+k^2\rho_0^4+\rho_0^2\rho_0'^2+\pa{1-k^2\rho_0^4+\rho_0^2\rho_0'^2}\cos\pa{2k\eta_2}+2k\rho_0^3\rho_0'\sin\pa{2k\eta_2}}{2\eta^2k^4\rho_0^2}.
\ee
In contrast, for modes well outside the horizon, we obtain 
\be{rho2out}
\rho^2_{\rm out}=\pa{\frac{\eta_2}{\eta}}^2\frac{4\rho_0^4\pa{1-\eta_2\frac{\rho_0'}{\rho_0}}+\eta_2^2\pa{1+\rho_0^2\rho_0'^2}}{9\rho_0^2}.
\ee

\section{Comparison with observations}
To conclude, we illustrate the potential consequences of the model. Solving the Pinney equation is a time consuming task, since it requires very high numerical accuracy in the presence of the quantum corrections. It is not possible, or at least not sufficiently straightforward, to span diverse orders of magnitudes in the wave number $k$, and subsequently plot the precise features originated by the QG corrections, covering different choices of the parameters. We therefore limit our analysis to a single case and to a simplified expression for $m(t)$.

Let us first note that the phase in the gravitational wave function is given by 
\be{expini}
\re^{2i p(y-y_1)}\simeq\re^{-4i \frac{\M^2}{H^2}\pa{\frac{1}{\eta^3}-\frac{1}{\eta_1^3}}}\,.
\ee
This unusual expression is explained by the fact that a quantity with dimensions of a volume has been ``hidden'' by setting $L=1$ after (\ref{act}). If this quantity (we shall rename it $\bar k^{-1}$, for simplicity) is made explicit again\footnote{This can be done by replacing $\eta\rightarrow  \eta\, \bar k$, $a\rightarrow a /\bar k$, and $k\rightarrow k/\bar k$.}, the ``gravitational phase'' (\ref{expini}) takes the form 
\be{grphase}
\re^{2ip(y-y_1)}\equiv\exp\paq{-i\alpha_{\rm P}\pa{\frac{1}{\eta^{3}}-\frac{1}{\eta_1^{3}}}}\;{\rm with}\; \alpha_{\rm P}=\frac{4 \M^2}{H_*^2\bar k^{3}}\,,
\ee 
where $\bar k$ is a quantity with the same dimensions as the wave number $k$. Once $\bar k$ is re-introduced, one may either assign to the scale factor dimensions of length (whereas the wave number and conformal time are dimensionless), or one can consider a dimensionless scale factor and assign dimensions of inverse length to the wave number and length to the conformal time.

In order to study the possible observable effects, it is important to note that the de Sitter evolution analyzed here must be considered as the leading order approximation to the slow-rolling inflationary dynamics, in which the constant potential acquires a small tilt that leads to the observed amplitudes and spectral indices for scalar and tensor perturbations (neglecting the kinetic energy of the inflaton). Moreover, as already mentioned, we shall limit our analysis to one particular case which does not require an unfeasible numerical effort, and we simplify the analytical expressions by keeping only first order terms in $\epsilon$ in the ``mass'' (\ref{mdef}).

Let us recall that the observable CMB window is the interval
\be{CMBwin}
\frac{k}{a_0}\in[10^{-4},10^{-1}]\,{\rm Mpc^{-1}}\;{\rm and\;we\; define}\; \frac{k_L}{a_0}=10^{-4}{\rm Mpc^{-1}},\; \frac{k_s}{a_0}=10^{-1}{\rm Mpc^{-1}}\,,
\ee
where $a_0$ is the scale factor today (which is generally fixed to one), and the interval spans three orders of magnitude. The largest scale corresponding to the CMB modes which re-enter the horizon today is $k_L$ with
\be{retod}
\frac{k_L}{a_0}\sim H_0\sim 1.4\cdot 10^{-42}\,{\rm GeV}. 
\ee
The 2018 Planck estimate of $H_*/\M$ for the vanilla $\Lambda$CDM + single field inflation sets the following upper limit on the tensor to scalar ration $r$:
\be{uplimV}
\frac{V_*}{\M^4}\equiv \frac{3\pi^2 A_s}{2}r<\frac{\pa{1.6 \cdot 10^{16}\,{\rm GeV}}^4}{\M^4} \;(95\%\;{\rm confidence\; level})\,,
\ee
which is derived from the (non-)observation of tensor modes, and $A_s$ is the amplitude of scalar perturbations, or equivalently
\be{uplim}
\frac{H_*}{\M}\sim 10^{-4} r^{1/2}<2.5 \cdot 10^{-5} \;(95\%\;{\rm confidence\; level}).
\ee
Henceforth, we restrict the parameter space and set the tensor to scalar ratio to $r=10^{-4}$, which satisfies the constraints above with values of $r$ that are not too small (and thus below the sensitivity of next-generation CMB surveys).

Let us note that given the constraint (\ref{uplimV}) and the range of validity of the approximation (\ref{compapp}), one has
\be{alargeapp}
\lambda^{1/6}\sim6\cdot 10^{-2} \,r^{1/6}\;{\rm and}\;a\gg \frac{\bar k}{\Mt \,\lambda^{1/6}}\simeq \frac{10^2\bar k}{6^{3/2} \,r^{1/6}}\M^{-1}\simeq 3\cdot 10^1\bar k \,\M^{-1}.
\ee
Therefore, when $a$ is of order $\sim \bar k\cdot10^2\,\M^{-1}$, the plane wave approximation (\ref{solgr}) is acceptable and classical time can be defined. During inflation, the modes in the interval (\ref{CMBwin}) exit the horizon at different times $\eta_k=-k^{-1}$, and, in particular, $k_L$ exit the horizon first (at $\eta_L$). Given the approximate scale factor evolution $a(\eta)\simeq -\pa{H_*\eta}^{-1}$, at $a\sim a_{\rm P}\equiv 100\,\bar k\,\M^{-1}$, when classical time arises, one finds 
\be{etaP}
\eta_{\rm P}=-\frac{\M}{10^2\,\bar k\,H_*}\sim -\frac{10^6}{10^2\,\bar k}\sim -\frac{10^4}{\bar k}.
\ee
If, for example, $\bar k=10^2$, then $\eta_P=-10^2$, and if we assume that $k_L$ and $k_s$ are still well inside the horizon at $\eta_P$ ($-k_{L,s}\eta_P\gg 1$), then we may set $k_L\simeq 1$.

In the following example, we take $\bar k=10^3$ in order to shorten the numerical calculations. As a consequence, given the choice of $r$, we have $\alpha_P=4\cdot 10^3$ and $\eta_P\sim -10$. We consider the resonant transmission occurring in the interval $[\eta_1,\eta_2]$ with $\eta_1=-10\sim \eta_P$ and $\eta_2\simeq -7.3$, so that a single oscillation of the gravitational wave function occurs in this interval (therefore minimizing the QG effects). In contrast, we take $\epsilon=10^{-1}$, which is quite large, and we span the 3 orders of magnitude interval $k\in2\cdot[10^{-1}, 10^2]$ in the numerical simulation. The longest wavelength modes in this interval are very close to the horizon exit, and, therefore, the condition $-k\eta\gg 1$ is not properly satisfied for them. However, in order to illustrate the possible effects, it is worth studying the evolution of these modes as well.  

\subsection{Numerical simulations}
The qualitative behavior of the solutions of the Pinney equation in the regime we are considering ($-k\eta\gg 1$) can be described as follows. First, the Pinney equation may be conveniently rewritten in terms of $y\equiv \sqrt{m}\rho$ as
\be{Pin2}
y''+\pa{\frac{\omega^2}{m^2}+\frac{1}{4}\frac{m'^2}{m^2}-\frac{1}{2}\frac{m''}{m}}y=\frac{1}{y^3}.
\ee
The initial conditions for $y$, at $\eta_1$, are related to those of the Pinney variable $\rho$ by
\be{yini}
y(\eta_1)=\sqrt{m(\eta_1)}\rho_{\rm dS}(\eta_1),\; y'(\eta_1)=\pa{\frac{1}{2}\frac{m'(\eta_1)}{m(\eta_1)}+\frac{\rho'_{\rm dS}(\eta_1)}{\rho_{\rm dS}(\eta_1)}}y(\eta_1)\,,
\ee 
and the solutions of (\ref{Pin2}) can be written with an expression analogous to (\ref{rhosol}), in terms of the solutions of the homogeneous equations associated with it. In the regime $-k\eta\gg 1$, we have $\omega^2\simeq k^2$ and two approximate (and independent) solutions are 
\be{indepsoly}
y_{\rm hom}^{(1,2)}\simeq \sqrt{m(\eta)}\re^{\pm i\int_{\eta_1}^\eta \frac{k}{m}\rd \eta}\,,
\ee
where, due to the oscillatory (complex) behavior of $m(t)$, the integral $\int^\eta \frac{1}{m}\rd \eta=r(\eta)+i s(\eta)$ and $r(\eta)$ and $s(\eta)$ are both positive. In particular, $r\sim (\eta-\eta_1)-\mathcal{O}(\epsilon)$ since the real part of $1/m$ oscillates around $1$. In contrast, its imaginary part oscillates around zero and the area subtended by the positive branch is larger than the area corresponding to the negative branch. Therefore, the imaginary contribution is positive as well. The sign of the imaginary contribution plays a significant role, especially for $k$ large, dampening the solution $y_{\rm hom}^{(1)}$ and simultaneously amplifying $y_{\rm hom}^{(2)}$. Then, for $k$ large enough (depending on $\epsilon$), the dampened contribution is subleading and
\be{largeksol}
\rho^2\simeq \re^{-2 i\int_{\eta_1}^\eta \frac{k}{m}\rd \eta}.
\ee 
The non-oscillating terms, originated by the product $y_{\rm hom}^{(1)}y_{\rm hom}^{(2)}$, and those proportional to $\sim(y_{\rm hom}^{(1)})^2$ can therefore be neglected. On the other hand, for smaller values of $k$, the product $y_{\rm hom}^{(1)}y_{\rm hom}^{(2)}$ remains the leading contribution and the oscillations around this ``slowly varying'' solution are very small.

\begin{widetext}
\begin{figure}[t!]
\centering
\includegraphics[width=6cm]{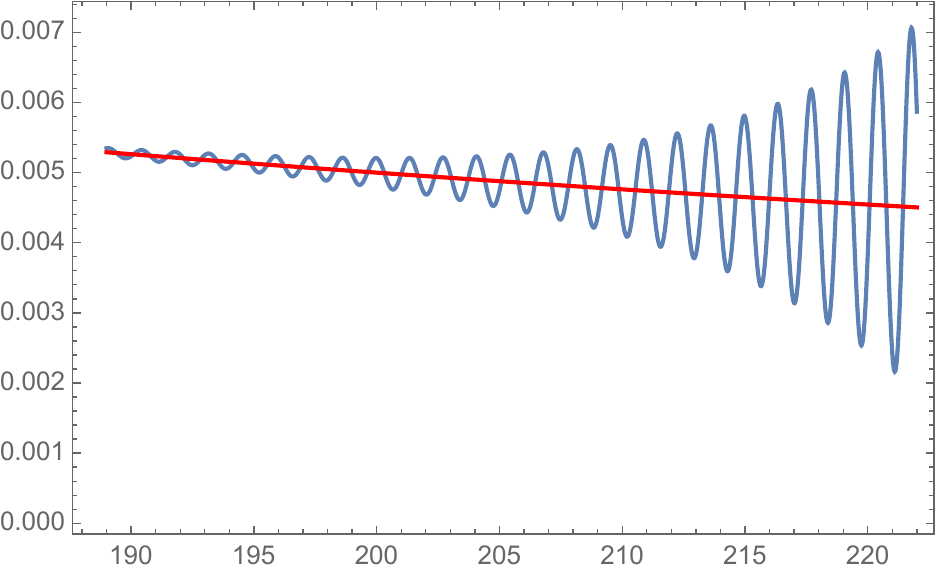}
\includegraphics[width=6cm]{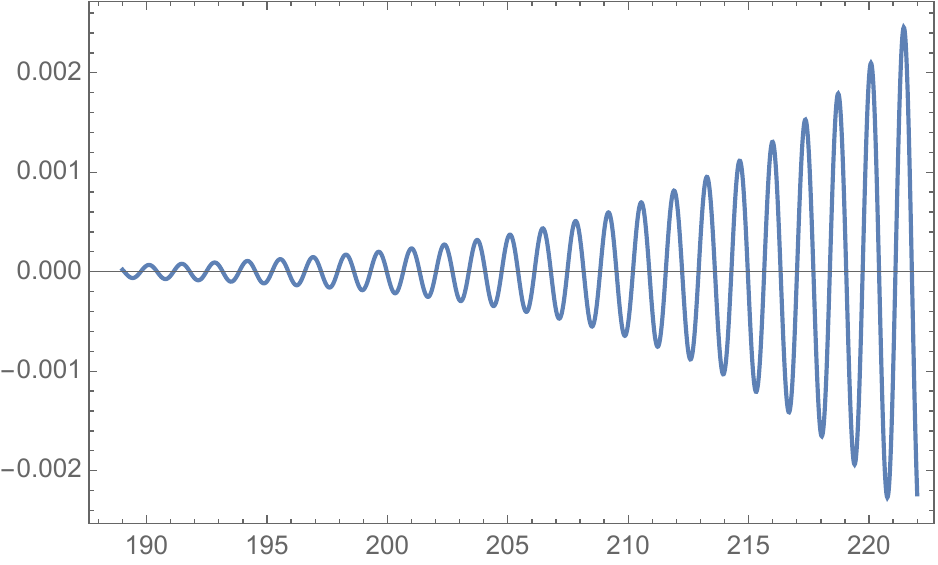}
\includegraphics[width=6cm]{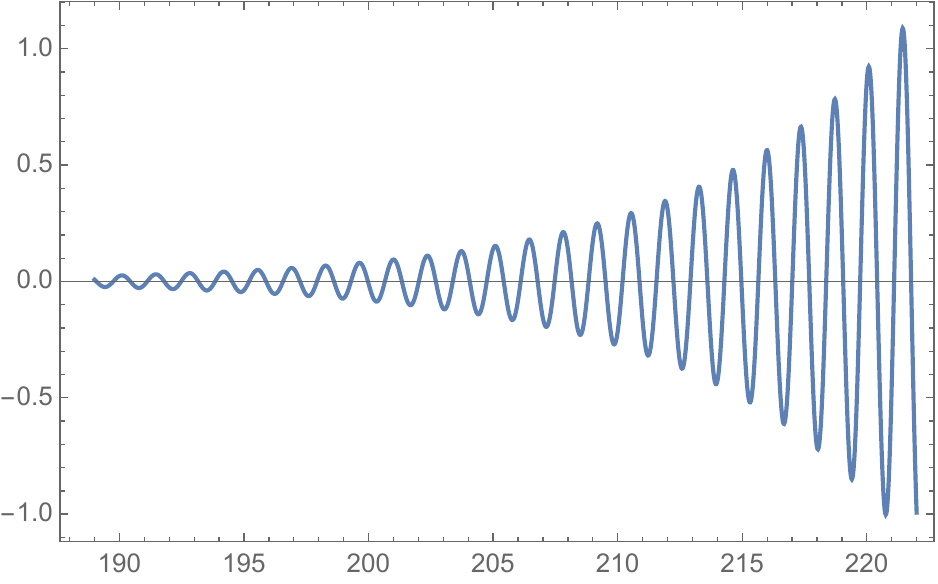}
\includegraphics[width=6cm]{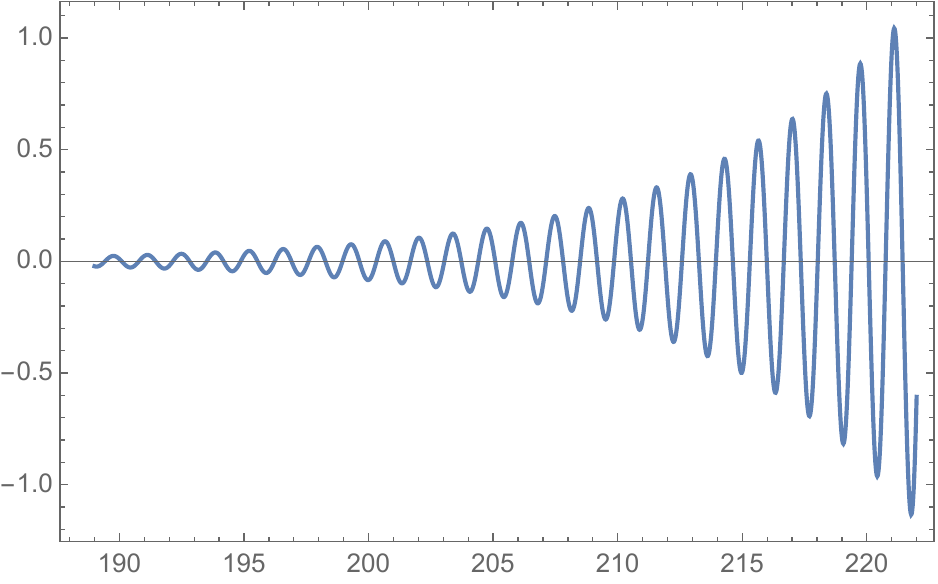}
\caption{\it The real and imaginary parts of $\rho^2(\eta_2)$ are shown in the top two figures. The red line is $\rho^2_{\rm dS}(\eta_2)$ (unperturbed). In the two figures below, the real and imaginary parts of $2\rho(\eta_2)\rho'(\eta_2)$ are plotted. We observe that $k\paq{\rho^2(\eta_2)-\rho^2_{\rm dS}(\eta_2)}\sim 2\rho(\eta_2)\rho'(\eta_2)$, and, for modes $-k\eta_2\gg 1$, then $\rho^2_{\rm dS}\simeq 1/k$.  Note that the perturbed solution oscillates around the unperturbed one. Moreover, the difference of a phase $-i$ between the two quantities is present: $-i \rho^2(\eta_2)\sim 2\rho(\eta_2)\rho'(\eta_2)$.
\label{f2}}
\end{figure}
\end{widetext}

\begin{figure}[t!]
\centering
\includegraphics[width=9.5cm]{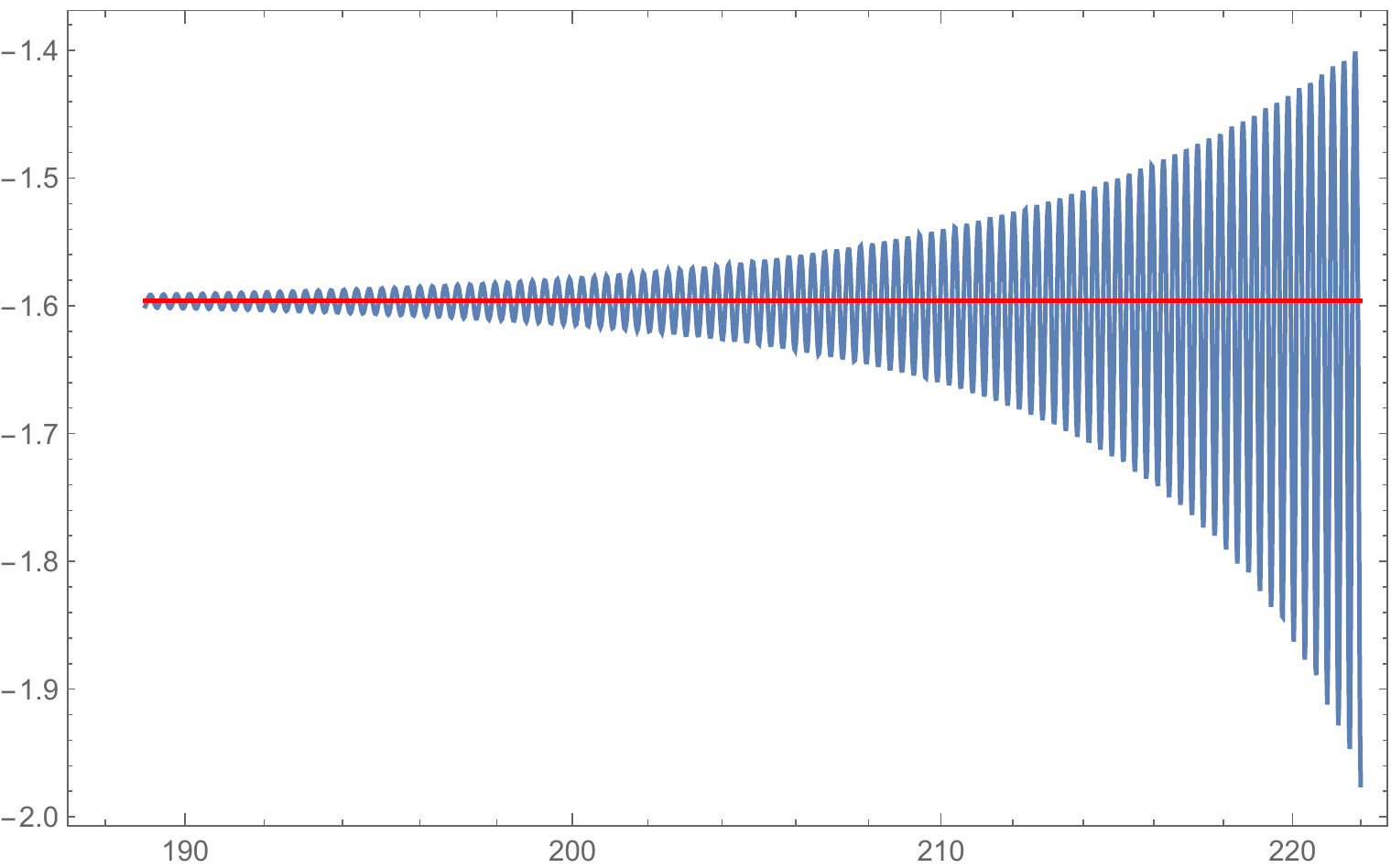}
\caption{\it The wave-number $k$ is plotted on the horizontal axis. On the vertical axis, $\log_{10}\Delta_s^2\eta^2$ for de Sitter (red line) and with the QG effects. 
\label{f3}}
\end{figure}

\begin{figure}[t!]
\centering
\includegraphics[width=9.5cm]{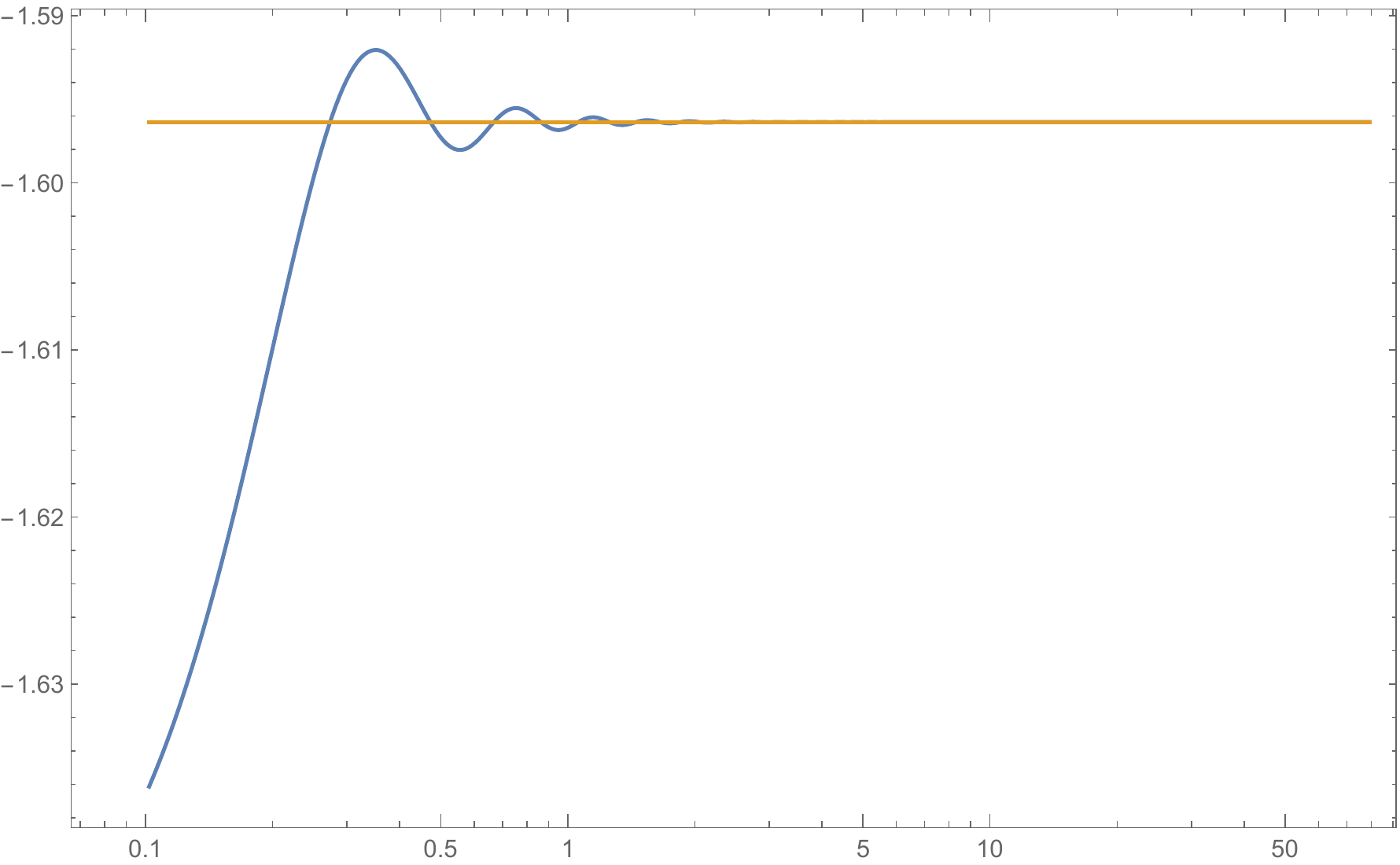}
\caption{\it The $\log_{10}$ of the wave-number $k$ is plotted on the horizontal axis. On the vertical axis, $\log_{10}\widetilde{\Delta}_s^2\eta^2$ for de Sitter (yellow line) and with the QG effects.
\label{f1}}
\end{figure}

In Figure (\ref{f2}), the values of $\rho^2(\eta_2)$ as function of $k$ are shown. For the reasons discussed above, the deviation from the unperturbed de Sitter solution (in red) increases as $k$ gets larger. The oscillation is essentially given by the real part of $\int_{\eta_1}^{\eta_2} m^{-1}\rd \eta\sim 2.3+0.07\,i$ and its frequency is $\Omega\sim -2r(\eta_2)$. We obtained a high-precision numerical fit of the oscillations in these plots with the following expression
\be{rho2fit}
\rho_{\rm fit}^2(\eta_2)=\frac{1+k^2\eta_2^2}{k^3\eta_2^2}\pa{1+\re^{a+b \,k}\re^{i\pa{\Omega\,k+c}}}\,,
\ee
with $a\simeq -27.182$, $b\simeq 0.120$, $c\simeq -1.567$. The functional dependence of $k$ in the expression (\ref{rho2fit}) is straightforwardly derived from (\ref{largeksol}). Correspondingly, 
\be{drho2fit}
\paq{\rho_{\rm fit}^2(\eta_2)}'= -2ik\frac{1+k^2\eta_2^2}{k^3\eta_2^2}\re^{a+b \,k}\re^{i\pa{\Omega\,k+c}}\,,
\ee
where essentially the factor $-2ik$ comes from the derivative w.r.t. $\eta$ of (\ref{largeksol}). If we insert the analytical expressions (\ref{rho2fit}) and (\ref{drho2fit}) in (\ref{rho2in}), we can reproduce the final numerical result plotted in Figure (\ref{f3}), where the (leading) visible oscillations have a frequency equal to $\Omega_{\rm PS}\simeq -2\,\eta_2-\Omega\sim 19.2$ as a consequence of the factors $\re^{i\pa{\Omega\,k+c}}$ in (\ref{rho2fit}) and (\ref{drho2fit}), as well as of the oscillation of frequency $2\eta_2$ present in the expression (\ref{rho2in}). In Figure (\ref{f3}), the quantity $\log_{10}(\widetilde\Delta_s^2)$ is plotted in a modes' interval which is visibly affected by the QG effects (at $k\sim 2\cdot 10^2$). The effects get larger if $k$ is increased, and they are essentially represented by an oscillation with frequency $\Omega_{\rm PS}$. The functional dependence of this frequency on $\eta_2$ and $\Omega$ can be easily determined and predicted if the parameters of the model are varied. Contrariwise, the low-energy regime of the spectrum is plotted in Figure (\ref{f1}). 

An oscillation with a frequency $\sim -2\eta_2$ and a decreasing amplitude (at lower $k$) is present. For increasing $k$, the oscillation amplitude gets smaller and finally starts increasing as Figure (\ref{f3}) shows. Notice that the oscillation is the consequence of the factors $\re^{\pm2ik\eta_2}$ in (\ref{rho2in}). However, the breakdown of the approximation $-k\eta_2\gg 1$ for the modes in the interval $[10^{-1},1]$ makes the analytical predictions deviate from the numerical result. Qualitatively, a decreasing amplitude is expected as the $\omega^2$ decreases near horizon, as this ``pumps'' energy in the system. The oscillation freezes for $k\rightarrow 0$, and the decreasing amplitude plotted in the figure is indeed just a part of an oscillation with increasing amplitude. Therefore, the power loss illustrated may become a power enhancement in some lower energy (and in this case unobservable) part of the spectrum.

Can we observe these QG effects? Concerning the long wavelength regime, the presence of oscillations is known to be compatible with observations, especially given the large uncertainties associated with the cosmic variance in this part of the spectrum \cite{powerloss}. In the short wavelength regime, it has been recently argued that the presence of a superimposed oscillation in this interval of the spectrum would cure or alleviate some anomalies and tensions in current cosmological data. This oscillation must involve a certain interval of modes and have a certain amplitude and frequency in order to improve the data fit. In particular (see \cite{Planck,fit}), a frequency $\omega_{\rm fit}\sim 300\,{\rm Mpc}^{-1}$ can improve the fit with observations, especially when it is restricted to the modes interval $2[10^{-2},10^{-1}]{\rm Mpc}$.  While the oscillation we obtain does not have all these peculiar features, as its amplitude increases monotonically in $k$, it would still be worth studying if its presence, with a given frequency\footnote{For the current choice of parameters together with $k$ having inverse length dimensions, we find $\Omega_{\rm PS}\sim 4\cdot 10^4\,{\rm Mpc}$.} and initial amplitude, could also contribute to the reconciliation of the tensions and anomalies in the different datasets, or whether it could be simply an observable feature overlooked by the data analysis. It is noteworthy that error bars in CMB (Planck) data also increase in the large $k$ interval, in particular close to $k_s$. Even so, the amplitude of the oscillation may be large enough to be detectable but further analysis it is necessary in order to understand to what extent it can be observable in the CMB spectrum. A very high frequency oscillation could be unobservable, which would result in a coarse grained (averaged) effect.

\section{Conclusions}
We have studied the possible effects arising from the quantum superposition in the gravitational wave function in the context of the traditional BO approach to quantum gravity. For inflationary cosmologies in a homogeneous background, the Hamiltonian constraint (Wheeler-DeWitt equation) is a second-order PDE for the wave function of the Universe, which is akin to a time-independent Schr\"odinger equation. The WDW equation can be decomposed into an infinite set of coupled PDEs through a BO decomposition: the equation for the gravitational wave function (which becomes a second order differential equation) and the remaining system of equations describing the matter degrees of freedom (the Mukhanov-Sasaki equations plus the equation for the homogeneous inflaton). In this approach, time is not taken to be fundamental, but rather it can be introduced in the quantum equations through the ``probability flux'' of the gravitational wave function. If this function is approximately like a plane wave, equations for (quantum) matter take the traditional semiclassical form (apart from tiny non-adiabatic corrections). In contrast, if one considers a sum of plane waves having opposite probability flux, the matter equations also contain other QG corrections that result from this superposition, which can be a consequence of the initial conditions of the universe (possibly including a bounce \cite{bounce}), or it can be limited in time and originate from a sudden variation of the energy density driving the expansion of the universe.

Concretely, we considered the cases of a potential well and that of potential barrier.
In both cases, the equations for matter are modified but they can be formally solved. The resulting inflationary spectra are computable provided time can be defined, at least perturbatively. In fact, the mathematical approach we adopted in order to solve the modified matter equations is novel. Instead of using a perturbative approach, we solved the equations exactly (at least formally) expressing the exact solutions in terms of the Pinney variable, evolution of which can be found numerically, and thus beyond the perturbative regime.

Finally, we studied the possible observable consequences for a simplified case, in which all the general and qualitative physical features of the QG effects could be ascertained. These effects can modify both the long and the short wavelength part of the spectrum, and they superimpose an oscillation to the standard semiclassical spectra. The frequencies of the oscillation for $k$ small is different (and smaller) from that at large $k$'s. The spectrum is essentially unmodified in the intermediate region, which can be arbitrarily enlarged or shrunk on varying the model parameters, and this also affects the modulation derived from the QG effects.

It was recently argued that a modifications in the high frequency part of the spectrum may cure the anomalies that seem to be present in Planck data and reconcile the tensions among CMB observations and other astrophysical datasets (for example the well-known tension coming from the local measurement of the Hubble constant). We believe that it is worth analyzing whether the QG effects described in this article may have similar consequences. This would provide an interesting connection between a conservative approach to quantum cosmology and the detailed physics of the early Universe. We hope to address this further analysis in future work. 

\vspace{-0.5cm}
\begin{acknowledgements}
L.C. thanks the Dipartimento di Fisica e Astronomia of the Universit\`{a} di Bologna as well as the I.N.F.N. Sezione di Bologna for financial support.
\end{acknowledgements}

\end{document}